# Super-efficient temporal solitons in mutually coupled optical cavities

Xiaoxiao Xue[*], Xiaoping Zheng, and Bingkun Zhou

*Department of Electronic Engineering, Beijing National Research Center for Information Science and Technology, Tsinghua University, Beijing 100084, China*
[*xuexx@tsinghua.edu.cn](mailto:xuexx@tsinghua.edu.cn)

**Abstract**

A coherently driven Kerr optical cavity is able to convert a continuous-wave laser to a sequence of ultrashort soliton pulses, enabling the generation of broadband and mode-locked frequency combs. Kerr cavity solitons are balanced through an energy exchange with the driving pump field. Improving the energy conversion efficiency from the pump to the soliton is of great significance for practical applications, but remains an outstanding challenge due to a limited temporal overlap between the soliton and the pump. Here, we report the discovery of temporal Kerr solitons in mutually coupled cavities instead of a traditional single cavity. We propose a strategy for breaking the limitation of pump-to-soliton energy conversion, and connect the underlying mechanism to impedance matching in radiofrequency electronic circuits. With macro optical fiber ring cavities which share the same physical model as miniature optical microresonators, we demonstrate nearly one-order improvement of the efficiency. Our findings pave the way towards super-efficient soliton microcombs based on optical microresonators with ultra-high quality factors.

Dissipative solitons are localized particle-like structures which are double balanced by gain and loss, and nonlinearity and dispersion (or diffraction) [1], [2]. The study of dissipative solitons spreads in a large variety of different areas, has led to many exciting scientific findings and useful applications. Dissipative temporal solitons in coherently driven Kerr optical cavities have attracted great interest in recent years. The study arose in two different scenarios. One focused in the time domain, i.e., exciting and maintaining solitons to form optical buffers [3]; the other focused in the frequency domain, i.e., optical frequency comb generation with miniature microresonators [4]. The experimental demonstration of temporal cavity solitons was first performed in macro fiber ring cavities [3], and then repeated in optical microresonators [5], [6]. The researches from frequency and time domains then merged to a clear subject of cavity solitons [7], [8]. The discovery of microresonator solitons truly brings cavity solitons to a hot active research area because it enables integrated coherent frequency comb sources which may revolutionize many applications [9]-[17].

Following the historical trend, the researches of cavity solitons nowadays have been performed on mainly two platforms. One is macro fiber ring cavities [3], [18]-[21]; the other is miniature optical microcavities [5], [8], [22]-[26]. In comparison, fiber cavities are more convenient for precise frequency detuning control and can be easily investigated in real time due to their much slower time scale; microresonators are particularly useful as compact frequency comb generators for practical applications. Previous studies have shown that both platforms share the similar physical model and useful technical hints may be learned from each other.

The basic setup of Kerr comb generation is composed of a Kerr nonlinear resonator coherently pumped by a continuous-wave (CW) laser. In the frequency domain, the comb generation is explained by cascaded four-wave mixing which transfers energy from the pump to newly generated frequency lines. In the time domain, the comb mode locking is generally related to formation of temporal cavity solitons. By using resonators with high quality (Q) factors, significant intracavity power buildup can be achieved, resulting in very low pumping threshold for nonlinear four-wave mixing. Comb initiation with mW and sub-mW pump power levels has been demonstrated with various microresonator platforms [27]-[32]; and fully integrated soliton frequency comb has been reported [33]. Despite of a high efficiency in terms of the required pump power level, the energy conversion efficiency of soliton combs, which is defined as the ratio of the pump power converted to the other frequency lines, is generally rather low [34].

A Kerr soliton is a localized short pulse circulating in the cavity atop a CW pump background. The propagation loss of the soliton is balanced through an energy exchange with the pump when they coincide in time. The energy conversion efficiency is thus primarily limited by the duty cycle of the soliton pulse (i.e., ratio of the pulse width over the round-trip time of the cavity) [35]. In typical single-soliton comb generation, only a few percent (in many cases even less than 1%) of the pump power is transferred to the frequency comb. Several approaches have been proposed to increase the efficiency, such as exploiting dark solitons in a normal-dispersion region [35], [36], maintaining multiple solitons in the cavity [24], synchronously pulsed pumping [25], or synchronizing and combining multiple solitons [26]. Nevertheless, significant improvement of the conversion efficiency is still quite challenging due to degraded comb spectral flatness and increased complexity of the pump source.

In this article, we show that efficient conversion of the pump to the soliton comb can be achieved by exploiting a mechanism that mimics impedance matching in radiofrequency circuits. Instead of a traditional single cavity, we investigate soliton generation in dual cavities that are mutually coupled to each other. With macro fiber ring cavities which can be accurately investigated and modeled, we experimentally demonstrate coupled-cavity solitons whose efficiency is improved by nearly one order. A theoretical model based on a set of fully-mapped coupled nonlinear Schrödinger equations is developed. The numerical simulation results show excellent agreement with our experiments. The same strategy and physics can be applied to integrated optical microcavities with ultra-high Q factors, making it possible to achieve super-efficient soliton microcomb sources with conversion efficiency close to 100%.

**Efficient energy transfer with coupled modes**

In the frequency domain, the energy conversion efficiency can be explained from another point of view. It has been proved both theoretically and experimentally that bright temporal solitons exist in an effectively red-detuned region (i.e., the frequency (wavelength) of the pump laser field is lower (longer) than the resonant frequency (wavelength) of the cavity) [5]. This pump-resonator frequency detuning is critical to maintain the solitons but at the same time prevents the pump power efficiently entering the resonator. A prerequisite of high conversion efficiency is signifi-cant depletion of the pump after interacting with the resonator. However, most soliton Kerr combs show spectra with a high contrast between the pump and the comb lines, implying very poor conversion efficiency. The problem can simply be explained by the resonant conditions illustration in Fig. 1. When the pump frequency matches that of the resonator (cavity 1) and the coupling rate equals the decay rate (see Fig. 1a, case I), the pump power from port P can be completely transferred to the drop channel (port D) – a condition usually called critical coupling. Here the meaning of the drop channel is generalized which takes into account the intrinsic loss, the coupling loss related to a drop waveguide, and the nonlinear loss due to energy transfer from the pump to the comb. To achieve a high pump-to-comb conversion efficiency, the nonlinear loss should be far beyond the other losses (see Supplementary Section 4 for more discussions). The second panel (II) of Fig. 1a shows the case when there is a frequency detuning between the pump and the resonator. Complete power transfer to the drop channel is impossible. This is generally the case in soliton comb generation. But if there is another resonator (denoted as cavity 2) coupled with cavity 1 and the input field (for simplicity, cavity 2 is assumed intrinsically lossless), full power transfer to the drop channel of cavity 1 can be achieved when the following condition is satisfied (see Fig. 1a, case III; and Supplementary Section 2 for the mathematic details).

$$\kappa_P = \frac{|\kappa_C|^2 \kappa_D}{(\omega_P - \omega_1)^2 + \kappa_D^2} \quad (1)$$

$$\omega_2 = \omega_P - \frac{|\kappa_C|^2 (\omega_P - \omega_1)}{(\omega_P - \omega_1)^2 + \kappa_D^2} \quad (2)$$

where $\omega_1$, $\omega_2$, and $\omega_P$ are the angular frequencies of the resonators and the pump field; $\kappa_D$ is the coupling rate between cavity 1 and the drop channel; $\kappa_C$ is the coupling rate between cavity 1 and cavity 2; $\kappa_P$ is the coupling rate between cavity 2 and the pump field.

The principle of maximizing the energy transfer here resembles the classical impedance matching in radiofrequency electronic circuits illustrated in Fig. 1b. A single resonator can be mapped to a complex load connected to a transmission line, while dual-coupled resonators can be mapped to cascaded LC networks. The transmission coefficient of the optical field from port P to port T in Fig. 1a shares the similar mathematical formula as the voltage reflection coefficient of the circuits in Fig. 1b, with the following relations $\kappa_P \leftrightarrow 1/(2Z_m C_2)$, $\kappa_C \leftrightarrow 1/(2Z_0 \sqrt{C_1 C_2})$, $\kappa_D \leftrightarrow 1/(2R_L C_1)$, $\omega_1 \leftrightarrow 1/\sqrt{L_1 C_1}$, and $\omega_2 \leftrightarrow 1/\sqrt{L_2 C_2}$ (see Supplementary Sections 1 and 2 for more details). Here, $Z_m$ and $Z_0$ are characteristic impedance of the transmission lines which are assumed lossless; $L_1$ and $L_2$ are inductance; $C_1$ and $C_2$ are capacitance; and $R_L$ is load resistance. Mapping optical resonators to electronic circuits has been employed as a useful tool in the synthesis of optical filters [37]. Here we introduce this methodology to resonator enhanced nonlinear optics for efficient energy transfer.

Inspired by this insight, a configuration of mutually coupled resonators for high-efficiency soliton Kerr comb generation is proposed in Fig. 1c. When the device is operating, a single temporal soliton is formed in cavity 1 while cavity 2 acts as a coupling adapter that matches the pump to cavity 1 to maximize energy conversion. The soliton comb is extracted from a drop port of cavity 1. At the through port of cavity 2, the pump field is ideally significantly degraded due to efficient power transfer to the comb. The nonlinearity of cavity 2 is not necessary in principle, but since both resonators are most likely fabricated with the same material in practical implementation, we will thus consider all their nonlinearities in the next. From another point of view, the role of cavity 2 can be regarded as a pump recycler that recycles the bypassing pump in single-pass single-cavity comb generation. A related idea of using an auxiliary cavity to improve power efficiency can be found in the advanced laser interferometer gravitational-wave observatory (LIGO) which has made the first successful detection of gravitational waves from a binary black hole merger [38].

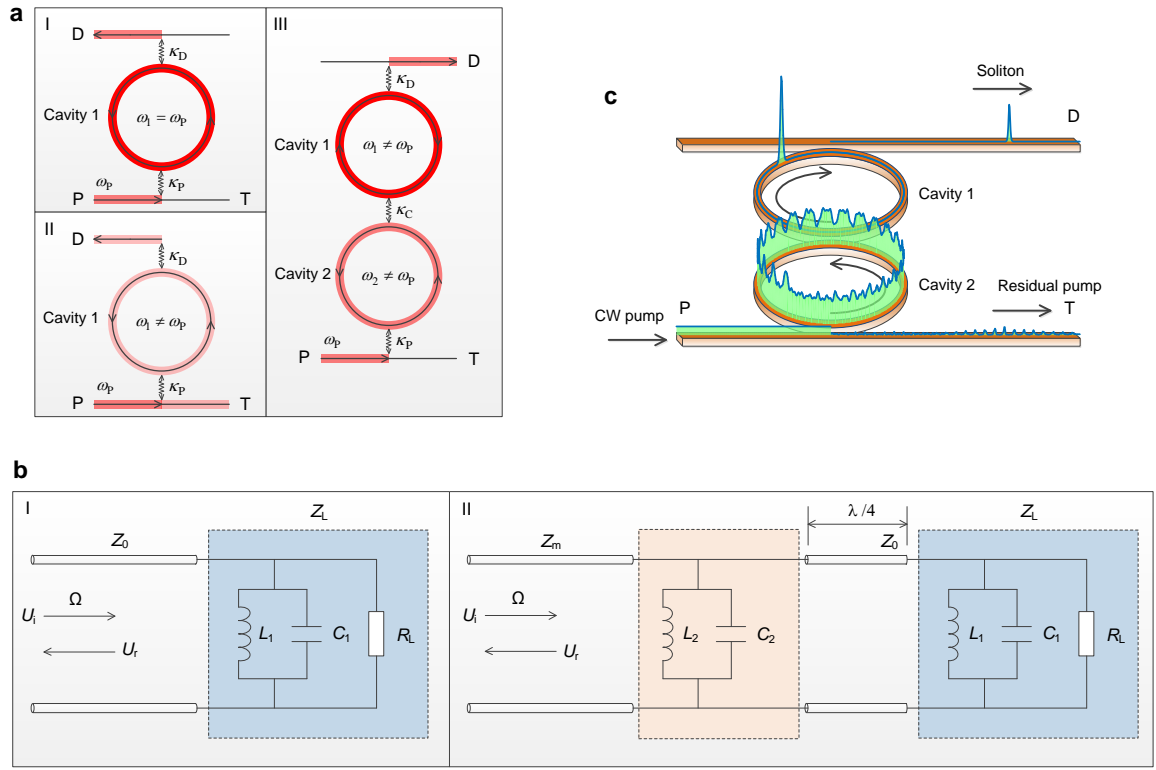

Figure 1 | Efficient energy transfer with coupled modes. (**a**) Optical cavities. Case I: when the pump frequency matches the resonator (cavity 1) frequency ($\omega_1 = \omega_P$) and the coupling rates satisfy $\kappa_P = \kappa_D$, all the power from port P can be transferred to port D; case II: when the pump frequency does not match the resonator ($\omega_1 \neq \omega_P$), full power transfer is impossible; case III: when a second resonator (cavity 2) is incorporated, full power transfer can be achieved again if its frequency and the coupling rates satisfy Eqs. (1) and (2). (**b**) Radiofrequency circuits that share the same mathematical model with the optical cavities. Case I: a complex load containing inductance ($L_1$), capacitance ($C_1$) and resistance ($R_L$) connected to a transmission line. Reflection arises when the signal frequency does not match the LC resonant frequency ($\Omega \neq 1/\sqrt{L_1 C_1}$) and the resistance does not match the characteristic impedance of the transmission line ($R_L \neq Z_0$). Case II: a parallel LC network ($L_2$ and $C_2$) is inserted quarter wave apart from the load, and another transmission line with characteristic impedance $Z_m$ is incorporated to eliminate the reflection. (**c**) Super-efficient soliton Kerr comb generation with mutually coupled resonators. The soliton circulates in cavity 1, while cavity 2 acts as a coupling adapter to maximize the power transfer from the pump to the soliton.

## Experimental results of mutually coupled optical fiber cavities

To demonstrate the idea, we performed experiments with mutually coupled fiber ring cavities. As explained in the introduction, macro fiber rings give us the convenience to precisely control the frequency detuning between the pump and the cavities, and to investigate the time-domain characteristics with low-speed electrical instruments. Moreover, hard excitation of solitons can be easily achieved with an external pulsed laser as in previous studies [3]. This gives us the ability to deterministically generate and manipulate the solitons.

A prototype with dual cavities is built with telecommunication single-mode optical fibers with a nonlinear Kerr coefficient $\gamma = 1.2 \text{ km}^{-1}\text{W}^{-1}$. The experimental setup is illustrated in Fig. 2. Cavity 2 has a round-trip length of 1.8 m corresponding to a free spectral range (FSR) of 113.8 MHz. The round-trip power loss is 18.7% (taking into account the coupling with cavity 1) corresponding to a finesse of 30. Cavity 2 is coupled to the pump through an 89/11 coupler. The estimated pump power threshold for modulational instability to occur in cavity 2 is 4.5 W (corresponding to an intracavity power of 48 W), well beyond the pump level in our experiments. Therefore, cavity 2 can be regarded as a linear cavity with negligible nonlinearity. Cavity 1 is much larger with a total round-trip fiber length of 75.3 m corresponding to a FSR of 2.7 MHz. The round-trip power loss is 23.1% corresponding to a finesse of 24. The two cavities are coupled to each other through a 95/5 wideband fiber coupler. A polarization controller is used to align the principle polarization axes of cavity 1 to cavity 2; and an isolator is inserted in cavity 1 to suppress the buildup of backward stimulated Brillion scattering. Suppose cavity 1 is being pumped through the 95/5 coupler, the estimated MI threshold is 0.58 W; and the minimum pump power for getting cavity 1 into a bistable region where solitons may exist is about 0.9 W.

To mitigate the influence of temperature fluctuations and mechanical vibrations, the round-trip lengths of the two cavities are feedback controlled with piezo-electric fiber stretchers. To stabilize cavity 2, a small portion of the pump light is sent through the 89/11 coupler in backward direction. The transmitted power is then stabilized to a specific value with a proportional–integral–derivative (PID) controller. Figure 2b shows the transmission curve plus bias measured by sweeping the voltage applied on fiber stretcher 2. Since the counter-clockwise propagating light in cavity 1 is blocked by an isolator (see Fig. 2a), only the response of cavity 2 itself is probed in this way with no coupling effect from cavity 1. The frequency detuning of the pump laser with respect to cavity 2 can be deduced from the zero-crossing position in Fig. 2b based on transmission reciprocity (recall that cavity 2 has negligible Kerr nonlinearity in our configuration). To stabilize cavity 1, the forward pump light is phase modulated by a radiofrequency signal and the Pound–Drever–Hall (PDH) signal is detected. Figure 2c shows the responses of the transmitted optical power and the PDH signal measured by sweeping the voltage applied on fiber stretcher 1 after cavity 2 is stabilized. A resonant dip can be observed which is a hybrid mode generated due to the coupling between cavity 1 and cavity 2. Cavity 1 is then stabilized by locking the PDH signal to zero level which coincides with the location with minimum power transmission (corresponding to maximum pump power injection into cavity 2).

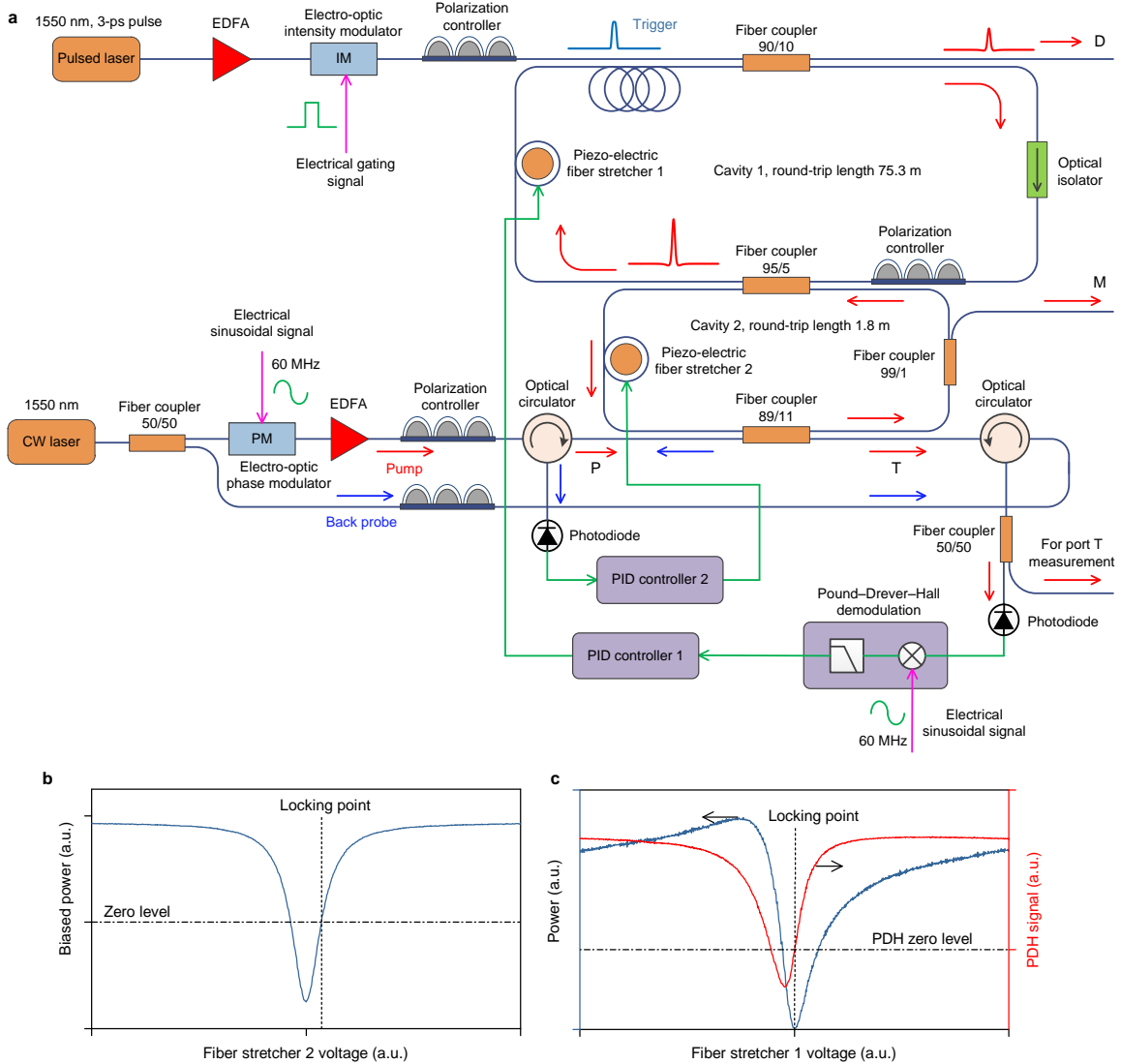

Figure 2 | Dual-coupled optical fiber ring cavities for studying the soliton dynamics. (**a**) Experimental setup. Cavity 1 has a total round-trip fiber length of 75.3 m and cavity 2 has a round-trip length of 1.8 m. Cavity 1 can support temporal Kerr solitons while cavity 2 can be regarded as a linear cavity with negligible Kerr non-linearity. The two cavities are respectively stabilized by locking the forward Pound–Drever–Hall (PDH) signal and the backward light power with proportional–integral–derivative (PID) controllers. Cavity 2 is pumped by a continuous-wave (CW) laser amplified with an erbium-doped fiber amplifier (EDFA). Temporal Kerr solitons can be excited by injecting a single-shot pulse into cavity 1. (**b**) Backward power transmission of cavity 2 measured by sweeping the voltage applied on fiber stretcher 2. (**c**) Forward power transmission of cavity 2 and PDH signal trace measured by sweeping the voltage applied on fiber stretcher 1 after cavity 2 is stabilized.

An erbium-doped fiber amplifier is used to increase the pump power to 0.27 W. To excite the soliton, a single-shot pulse is injected to cavity 1 through a 90/10 fiber coupler. The same coupler is also used to extract the soliton once it is generated. The full-width-at-half-maximum (FWHM) of the trigger pulse is 3 ps; and the peak power before entering cavity 1 is around 40 W. The pulse is shaped to third-order super-Gaussian to avoid serious distortions caused by self-phase modulation. Different from [3] in which the soliton is excited through cross-phase modulation between the pump and the trigger pulse, in our experiments, the trigger pulse has a central frequency roughly equal to that of the pump and directly evolves to a soliton after entering cavity 1 (similar to a method of applying intensity modulation to the pump laser [21]). After the soliton is formed, the output of the pulsed laser is then completely shut off to avoid any interference to soliton characterization. Figure 3a shows the optical spectrum corresponding to one single soliton in cavity 1, measured at the drop port (D) in Fig. 2. The numerical simulation result based on a set of coupled nonlinear Schrödinger equations is also shown (see Methods for detailed parameters and Supplementary Section 3 for equation derivation). The agreement between experiment and simulation is remarkable. The spectral envelop of the soliton follows a typical $sech^2$ function [5] with two weak spectral ears which in mode-locked fiber lasers are known as Kelly sidebands generated due to the discrete nature of the cavity [39].

The spectrum inside cavity 2 probed through a 99/1 coupler (i.e., port M) is shown in Fig. 3b. A strong pump line is observed, as well as a spectral pedestal similar to the soliton. The contrast between the pump and the pedestal is as high as 70 dB, implying that the optical field inside cavity 2 is mostly a CW pump. The power level of the pump line is 2.47 W – about 9.1-fold enhanced compared to that in the input fiber out of cavity 2 (i.e., port P). Note that if the CW laser directly pumps cavity 1 through the 95/5 coupler as in traditional soliton generation with a single cavity, the required pump power would be around 2.47 W to support a similar soliton as in Fig. 3a. In comparison, the pump power for the dual coupled cavities here is merely 0.27 W. The efficiency is thus improved by nearly one order. Actually, if cavity 1 is directly pumped, a pump power of 0.27 W cannot even get cavity 1 to a bistable region which is a prerequisite for solitons to exist. Therefore, the dual coupled cavities can significantly reduce the pump power for generating solitons. The soliton-like spectral pedestal in cavity 2 is generated due to mode coupling between the cavities in the mode-crossing regions. Limited by the resolution of our spectrum analyzer, the fine spectral features cannot be resolved. The pedestal actually contains periodic peak clusters as can be deduced from the electrical spectrum of the comb beat notes shown in the following Fig. 4b (see Supplementary Section 6 for simulated high-resolution optical spectra).

Figure 3c shows the spectrum at the through port (T) in Fig. 2. The pump power is reduced by 21 dB compared to the input power at port P. More than 99% of the pump power is transferred to cavity 2. Due to the high intrinsic loss of the fiber cavities arising from fiber splicing and the fiber components (such as isolators, fiber stretchers, etc.), about 68% of the pump power is dissipated by the intrinsic loss. Also, because of a relatively large cavity dimension corresponding to an extremely low soliton duty cycle ($6.2 \times 10^{-6}$), the resulting conversion efficiency of one single soliton is $4.6 \times 10^{-5}$ even after nearly one-order improvement. Therefore, fiber cavity solitons are investigated here as a more accessible physical equivalent of microresonator solitons rather than a usable comb source. Nevertheless, we note that such a distinct pump depletion, which is a prereq-uisite of high conversion efficiency, is impossible and has never been observed in traditional single cavity based single soliton generation due to a large effective frequency detuning between the cavity and the pump laser. For optical microresonators with ultra-high Q factors, the intrinsic loss may be greatly reduced such that the nonlinear loss related to comb generation becomes dominant. Outstanding conversion efficiency is thus achievable (see more discussions in the next section).

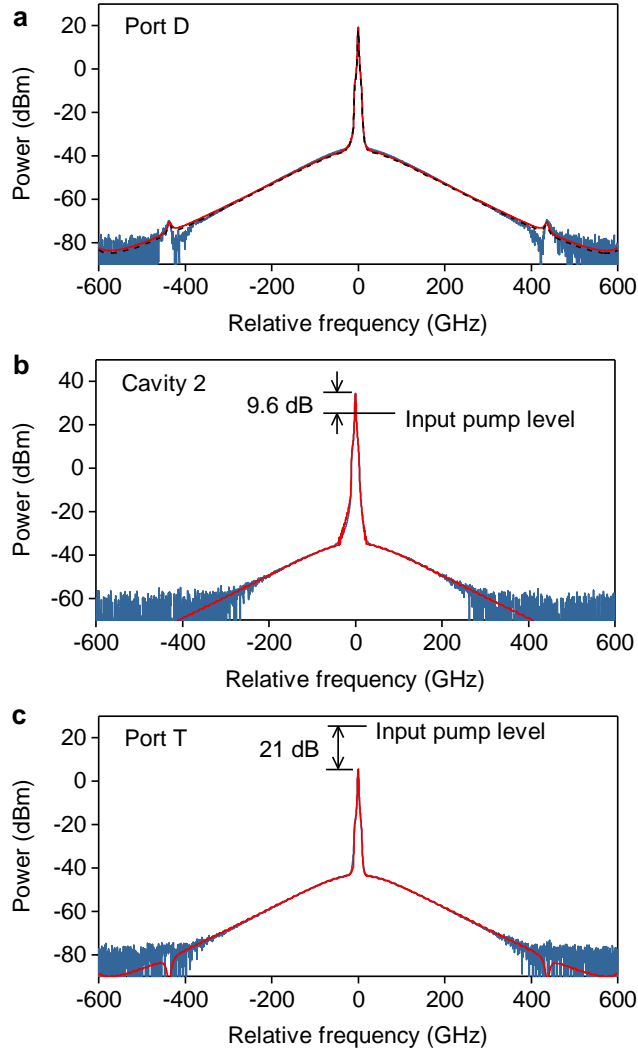

Figure 3 | Optical spectra (**a**) at port D, (**b**) inside cavity 2, and (**c**) at port T, when a single soliton is excited in cavity 1. The CW pump power at port P is 0.27 W. Dark blue: measured; red: simulated. Also shown in (**a**) is a simulated soliton spectrum (black dash) obtained by directly pumping cavity 1 through the 95/5 coupler with a significantly higher power (2.47 W). The filtering shape of the spectrum analyzer with a resolution of 2.5 GHz is taken into account in simulation. The pump line at port T is reduced by 21 dB compared to the input level, meaning that more than 99% of the pump power is transferred to cavity 2. The pump line in cavity 2 is enhanced by nearly one order compared to the input level. The dual coupled cavities can thus significantly reduce the external pump power for generating the solitons compared to one traditional single cavity.

Figure 4a shows the waveform at port D detected by a photodiode. Nice isolated pulses with a period of 0.37 μs can be observed. The waveform at port T is also shown. A band-stop fiber

Bragg grating is used to suppress the high pump line to increase the signal-to-noise ratio. The waveform at port T with the pump line suppressed contains a series of decaying pulses with a quasi-period of 8.8 ns (that is the round-trip time of cavity 2), in excellent agreement with the numerical simulation. The waveform inside cavity 2 is a strong pump with some decaying dips, as indicated by the simulation. The decaying dips are related to dispersive waves generated due to mode crossing between the two cavities. Figure 4b shows the electrical spectra of the comb beat notes at port D and port T. The spectrum at port D contains uniform lines in the measurement range, while the spectrum at port T is composed of periodic clusters corresponding to the fine features of the optical spectrum (see Supplementary Section 6). Figure 4c shows the measured and simulated autocorrelation of the soliton pulse from port D. The autocorrelation peak has a FWHM of 3.1 ps corresponding to a pulse width of 2.3 ps. Excellent agreement between measurement and simulation is obtained once again.

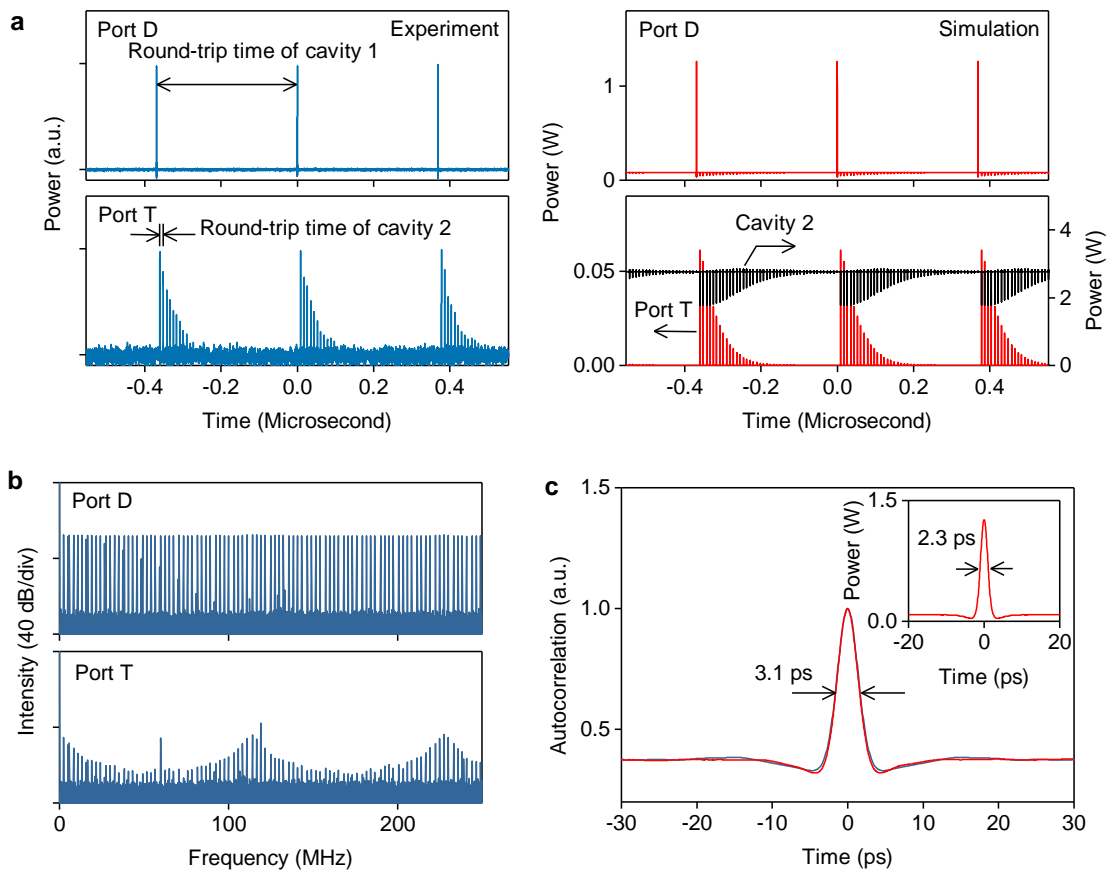

Figure 4 | Time-domain characterization of one single soliton in coupled fiber cavities. (**a**) Waveforms measured with a photodiode. A band-stop fiber Bragg grating is used to suppress the strong pump line to increase the signal-to-noise ratio. The waveform at port D contains periodic isolated pulses, while at port T contains decaying pulse trails. Note that the vertical scales for port D and port T are not identical. The pulse trails at port T are much weaker (more than one order) than the pulses at port D. Left: experiment; right: simulation. The simulated waveform inside cavity 2 is also shown (black), which is a strong pump plus some decaying dips. (**b**) Electrical spectra of the comb beat notes at port D and port T. (**c**) Autocorrelation of the soliton pulse from port D. Dark blue: measured; red: simulated. The inset shows the simulated soliton pulse in cavity 1.

**Computational results of intrinsically lossless microcavities**

For practical applications, soliton Kerr comb generation is generally performed with miniature optical microcavities which have compact volume and large FSRs. The intrinsic losses due to scattering and material absorption constitute one fundamental energy waste when the light circulates in the cavity. One continuing effort in microcavity fabrication is thus to reduce the intrinsic losses and increase the intrinsic Q factor. One interesting question is what conversion efficiency we can get if we have ideal resonators with completely no intrinsic losses (corresponding to infinite intrinsic Q). Note that for traditional comb generation with a single microresonator, the conversion efficiency of soliton combs is still quite limited even with an intrinsically lossless cavity [35]. Here we consider a similar case for mutually coupled microresonators and present numerical results based on fully-mapped coupled nonlinear Schrödinger equations. The accuracy of the theoretical model has been verified by the experimental results of fiber ring cavities presented above. The key parameters in our simulation, such as Kerr nonlinear coefficient and group velocity dispersion, are comparable to the typical values for silicon nitride microring resonators which can be fabricated with a complementary metal–oxide–semiconductor compatible technique [40]. Although intrinsically lossless microresonators do not exist in practice, the results presented in the next provide useful reference on the theoretical limit and may be approached with ultra-high-Q microresonators that are heavily over-coupled.

According to the principle explained above, only coupling of the pump between the two microresonators is necessary to facilitate energy transfer. To avoid strong interactions of the other modes which may destroy the soliton [41], slightly different radii are chosen for the two cavities. The FSRs of cavity 1 and cavity 2 are 200 GHz and 192.6 GHz, respectively. The group velocity dispersion of cavity 1 is anomalous ($-100$ $ps^2$/km) which is necessary to support bright temporal solitons. To suppress modulational instability in cavity 2 which is undesired in our scheme, the dispersion of cavity 2 is chosen to be normal (100 $ps^2$/km). The different signs of dispersion can easily be achieved by tailoring the waveguide dimensions in fabrication. The pump power injected from port P is 20 mW. The initial optical fields inside the microresonators are given by the steady-state CW solutions of the coupled equations. A single-shot Gaussian pulse, which has a FWHM of 0.2 ps and a peak power of 200 W, is introduced in cavity 1 to excite the soliton. The numerical results show that the pulse quickly evolves to a stable temporal soliton in nanoseconds. The soliton FWHM is around 24 fs. (See Methods for a detailed list of simulation parameters.)

Since strong depletion of the pump power is usually a clue of efficient energy conversion, we optimize the device parameters in our simulations to minimize the pump at port T. The optical fields in the microresonators and in the bus waveguides are shown in Fig. 5a. In the bus waveguides, the soliton at port D takes 98.0% of the total input power and the residual power at port T is only 2.0%. The optical field inside cavity 2 is very close to a CW pump with an average power of 1.9 W. Figure 5b shows the comb spectra in the bus waveguides. At port D, a smooth single-soliton comb can be observed. The conversion efficiency defined as the ratio of the comb power at port D excluding the pump line over the total input pump power is 94.2%. At port T, the pump line is significantly degraded which means the pump is close to effective critical coupling in comb operation. Similar to the case with fiber cavities, there are periodic comb clusters generated due to mode crossing between the two microresonators.

The field in cavity 1 closely resembles a soliton driven by a CW pump field of 1.9 W as in the case of a single microresonator. But in the case of single-resonator comb generation, the energy conversion efficiency is merely 1.0%. The mutually coupled microresonators can thus significantly improve the efficiency. Since the pump is almost completely depleted at port T, the (external) conversion efficiency here is close to the internal efficiency (96.1%) which is defined as the ratio of the comb power excluding the pump line over the total power in cavity 1 (see Supplementary Section 4 for more discussions). It has been shown that the internal efficiency of soliton combs can be relatively high [42], [43]. From another point of view, cavity 2 can be regarded as a pump amplifier that boosts the pump power for cavity 1 and a coupling adaptor that maintains a close-to-critical-coupling condition for the pump in comb generation.

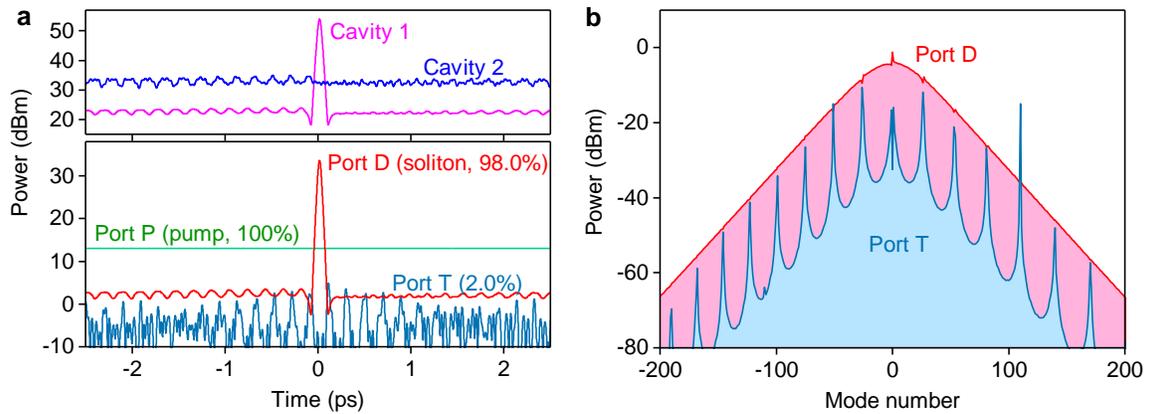

Figure 5 | Single-soliton Kerr frequency comb generation in mutually coupled optical microresonators assuming no intrinsic losses. Typical parameters of silicon nitride microring resonators are employed. The pump power is 20 mW. (**a**) Time-domain waveforms. Top: in the cavities; bottom: in the bus waveguides. The soliton at port D contains 98.0% of the total power, while the residual power at port T is merely 2.0%. (**b**) Comb spectra in the bus waveguides. Significant pump depletion can be observed at port T, corresponding to a high conversion efficiency of 94.2% (pump to comb-at-port-D).

## Summary and discussion

In summary, soliton Kerr comb generation with mutually coupled optical cavities is proposed for extraordinarily high energy conversion efficiency. The underlying mechanism resembles impedance matching in radiofrequency circuits. With macro fiber ring cavities, we demonstrated temporal Kerr solitons in mutually coupled resonators for the first time. A theoretical model based on a set of coupled nonlinear Schrödinger equations is developed, producing simulation results in excellent agreement with the experiments. For microresonators with no intrinsic losses, conversion efficiency close to 100% is expected. By considering practical microresonators with an intrinsic Q of $10^7$ (demonstrated with state-of-the-art silicon nitride integration technique [31], [32], [44]), conversion efficiency around 80% is achievable (see Supplementary Section 7 for more discussions). The super-efficient soliton comb will benefit a wide range of applications including biological and chemical sensing [10], high-capacity and long-haul fiber telecommunications [11], fast ranging lidar [13], [14], microwave photonic signal processing [16], [17], etc.

In our experiments and simulations, the soliton is excited by hard excitation with a single-shot pulse trigger. Numerical simulations show that solitons may also be spontaneously excited from modulational instability induced chaos by sweeping the resonator detuning and the pump power. Moreover, the mode coupling between the two resonators imposes a strong regulation on the soliton dynamics. Deterministic single soliton transition may be achieved, which is of significance for practical applications (see Supplementary Section 7).

Note that thermo-optic effect is not considered in our current numerical model. We did observe thermal effect of our fiber rings which was not revealed in previous studies. The cavity thermal drifting is relatively slow with a response time of sub-second (see Supplementary Section 5). The fiber stretchers and feedback circuits are fast enough to catch the cavity resonance in case of any intracavity power transitions. For microresonators with much more compact volume, the thermal drifting speed is much faster. Several technical tricks have been proposed to overcome the thermal instability related to intracavity power transitions, such as pump laser scanning [5], power kicking [45], and microresonator tuning [46]-[49]. For mutually coupled microresonators, independent tuning of each resonator is essential. Depending on different materials that form the microresonator, potential methods may include integrated thermal heaters [46], [48], free-carrier dispersion effect [47], and electro-optic effect [49]. On the other hand, the impact of thermal effect will be reduced by increasing the microresonator intrinsic Q (i.e., reducing the power absorbed which generates heat). The method we used to lock two fiber ring cavities may also be employed for microresonator control as long as a fast tuning approach is available.

The coupled nonlinear Schrödinger equations we developed can easily be modified to include higher-order terms such as Raman scattering, higher-order dispersion, and self-steepening. The impact of higher-order terms may become more significant for ultrashort soliton pulses with ultrabroad spectra (e.g., octave-spanning). Since our model is based on a full map, the equations can also be used to model super cavity solitons which exist in a large pump-resonator detuning region where a mean-field approach is no longer valid [20], [50]. Even higher conversion efficiency may be achieved by exploring new soliton states [43].

We note that coupled-microresonator structure was proposed before for tuning the local dispersion as well as the coupling condition [51], [52]. The investigations were not put in the context of high-efficiency temporal solitons and the demonstrated efficiency was still limited to a few percent. No corresponding theoretical model or experiments have ever been reported to provide insight into the soliton dynamics before the current contribution. Mutually coupled cavities provide more freedom of linear and nonlinear mode interactions, and show more rich dynamics than a single cavity. We expect our findings will have a wide impact on resonator enhanced nonlinear science.

## Acknowledgement

The authors would like to thank Andrew M. Weiner, Michael L. Gorodetsky, and François Leo for fruitful discussions; and Sigang Yang for lending the dispersion compensating fiber.

## Methods

**Stabilizing and locking dual coupled fiber ring cavities.** The cavities are built with telecom single-mode fibers (G.652 SMF). Cavity 1 has a round-trip length of 75.3 m, and cavity 2 has a round-trip length of 1.8 m. The two cavities are coupled to each other through a 95/5 wideband fiber coupler. The setup is put in a cabinet to reduce environment temperature fluctuations and vibrations. Piezo-electric fiber stretchers are used to further stabilize the cavity length. A rather short one (~0.4 m) is used in cavity 2, and a longer one (~5 m) is used in cavity 1. Two home-built analog proportional–integral–derivative (PID) controllers are used to generate the feedback signals. A wavelength-fixed laser (OEwaves) with a narrow linewidth of ~30 Hz is used as the pump source. The power is amplified by an erbium-doped fiber amplifier (EDFA), and coupled to cavity 2 through an 89/11 coupler. A fiber Bragg grating (FBG) with a bandwidth of ~10 GHz is used to suppress the amplified spontaneous emission noise from the EDFA. To stabilize cavity 2, a portion of the pump laser power is sent through the 89/11 coupler in backward direction. The transmitted power is then locked to a specific value. Since there is one isolator in cavity 1, the back-probed transmission of cavity 2 is not affected by cavity coupling. The detuning between the pump and cavity 2 can thus be deduced from the backward transmission (i.e., $\delta_2$ in the theoretical model). To stabilize cavity 1, the phase of the forward pump is modulated by a single tone (~60 MHz) and the Pound–Drever–Hall (PDH) signal is detected after transmission. The PDH signal is locked to zero corresponding to minimum power transmission in the forward direction. Numerical simulations are performed based on the experimental parameters to find the proper detuning between the pump and the cavities. The minimum forward transmission point corresponds to a bistable region of cavity 1 which can support cavity solitons. This should be the case in an optimized system. It should be noted that in general the minimum transmission point does not always correspond to a bistable region if the system is not optimized.

**Excitation and characterization of solitons.** A fiber mode-locked laser (MLL) which generates short pulses at a repetition rate of 50 MHz is used as a trigger to excite the solitons. The pulse is shaped to third-order super-Gaussian with a width of ~3 ps by using a commercial pulse shaper (Finisar) to avoid serious distortions caused by self-phase modulation. The pulse peak power is amplified to ~40 W with a method of chirped pulse amplification [53]. An optical intensity modulator (EOspace) is used as a fast optical switch to select the trigger pulses. The output of the MLL is completely shut off after the soliton is excited, to avoid interference to soliton characterization arising from a limited extinction ratio of the intensity modulator. To measure the waveform of the soliton pulses, an FBG with a bandwidth of ~10 GHz is used as a band-stop filter to suppress the high pump line to increase the signal-to-noise ratio of the electrical signal detected with a photo-diode. To measure the intensity autocorrelation of the soliton pulses, around 2300 solitons are excited in cavity 1 to increase the peak-to-background contrast of the autocorrelation trace. The fiber dispersion between the output of cavity 1 and the input of the autocorrelator is carefully compensated with a length of dispersion-compensating fiber.

**Numerical simulation.** The evolution of the optical fields is described by the following coupled nonlinear Schrödinger equations with a full map (see Supplementary Section 3 for equation derivation).

$$\frac{\partial E_1}{\partial z} = \left[ -\alpha_{i1} - i\delta_1 - i\frac{k_1''}{2}\frac{\partial^2}{\partial \tau^2} + i\gamma_1 |E_1|^2 \right] E_1 - \sum_{n=-\infty}^{+\infty} \delta(z-nL)\left(1-\sqrt{1-\theta_1}\right) E_1 \\ + \sum_{n=-\infty}^{+\infty} \delta\left(z-nL-\frac{L}{2}\right)\left[ i\sqrt{\theta_{12}} E_2 - \left(1-\sqrt{1-\theta_{12}}\right) E_1 \right] \quad (3)$$

$$\frac{\partial E_2}{\partial z} = \left[ -\alpha_{i2} - i\delta_2 - \Delta k' \frac{\partial}{\partial \tau} - i\frac{k_2''}{2}\frac{\partial^2}{\partial \tau^2} + i\gamma_2 |E_2|^2 \right] E_2 + \sum_{n=-\infty}^{+\infty} \delta(z-nL)\left[ i\sqrt{\theta_2} E_{\text{in}} - \left(1-\sqrt{1-\theta_2}\right) E_2 \right] \\ + \sum_{n=-\infty}^{+\infty} \delta\left(z-nL-\frac{L}{2}\right)\left[ i\sqrt{\theta_{21}} E_1 - \left(1-\sqrt{1-\theta_{21}}\right) E_2 \right] \quad (4)$$

$$E_{\text{T}} = i\sqrt{\theta_2}\, E_2\big|_{z=nL} + \sqrt{1-\theta_2}\, E_{\text{in}} \quad (5)$$

$$E_{\text{D}} = i\sqrt{\theta_1}\, E_1\big|_{z=nL} \quad (6)$$

where $E_1$, $E_2$, $E_{\text{T}}$, $E_{\text{D}}$ and $E_{\text{in}}$ are complex field amplitudes in cavity 1, cavity 2, port T, port D, and port P, normalized so that $|E_*|^2$ represents the power; $z$ is the propagation distance in the cavities (the dimension of cavity 2 is scaled such that the optical field in it shares the same circumferential coordinate as in cavity 1); $\alpha_{i1,i2}$ intrinsic amplitude loss per unit length; $\delta_{1,2}$ phase detuning per unit length; $\Delta k'$ group velocity mismatch; $k_{1,2}''$ group velocity dispersion; $\gamma_{1,2}$ nonlinear coefficients; $L$ circumference of cavity 1; $\theta_1$ power coupling ratio between cavity 1 and the drop channel; $\theta_2$ power coupling ratio between cavity 2 and the pump; $\theta_{12} = \theta_{21}$ power coupling ratio between cavity 1 and cavity 2; $\delta(\cdot)$ the Dirac function.

The equations are numerically integrated with the split-step Fourier method [54]. For the results of fiber ring cavities shown in Figs. 3 and 4, the parameters are $FSR_1 = 2.709$ MHz, $FSR_2 = 113.8$ MHz, $L = 75.27$ m, $\alpha_{i1} = 7.622\times10^{-4}$ m$^{-1}$, $\alpha_{i2} = 2.710\times10^{-4}$ m$^{-1}$, $k_1'' = -21.1$ ps$^2$km$^{-1}$, $k_2'' = -0.5023$ ps$^2$km$^{-1}$, $\gamma_1 = 1.2\times10^{-3}$ m$^{-1}$W$^{-1}$, $\gamma_2 = 2.857\times10^{-5}$ m$^{-1}$W$^{-1}$, $\Delta k' = -4.788\times10^{-9}$ s·m$^{-1}$, $\theta_1 = 9.864\times10^{-2}$, $\theta_2 = 0.1152$, $\theta_{12} = 4.298\times10^{-2}$, $E_{\text{in}} = 0.5196$ W$^{-1/2}$, $\delta_1 = 6.005\times10^{-3}$ m$^{-1}$, $\delta_2 = 1.355\times10^{-3}$ m$^{-1}$. The corresponding frequency detuning of cavity 1 ($\Delta f_1$) and cavity 2 ($\Delta f_2$) with respect to the pump is 0.1950 MHz and 1.847 MHz, respectively. The relation between $\Delta f_{1,2}$ and $\delta_{1,2}$ is given by $\Delta f_{1,2} = FSR_{1,2} \cdot \delta_{1,2} L/(2\pi)$.

For the results of microresonators shown in Fig. 5, the parameters are $FSR_1 = 200$ GHz, $FSR_2 = 192.6$ GHz, $L = 1.499$ mm, $\alpha_{i1} = \alpha_{i2} = 0$, $k_1'' = -100$ ps$^2$km$^{-1}$, $k_2'' = 100$ ps$^2$km$^{-1}$, $\gamma_1 = \gamma_2 = 1$ m$^{-1}$W$^{-1}$, $\Delta k' = 1.282\times10^{-10}$ s·m$^{-1}$, $\theta_1 = 9.114\times10^{-3}$, $\theta_2 = 1.063\times10^{-2}$, $\theta_{12} = 3.038\times10^{-3}$, $E_{in} = 0.1414 \text{ W}^{-1/2}$, $\delta_1 = 1.201 \times 10^2 \text{ m}^{-1}$, $\delta_2 = 8.809 \text{ m}^{-1}$. The corresponding frequency detuning of cavity 1 ($\Delta f_1$) and cavity 2 ($\Delta f_2$) with respect to the pump is $5.730 \text{ GHz}$ and $0.4048 \text{ GHz}$, respectively.

# Supplementary information to
# Super-efficient temporal solitons in mutually coupled optical cavities

Xiaoxiao Xue[*], Xiaoping Zheng, and Bingkun Zhou

*Department of Electronic Engineering, Beijing National Research Center for Information Science and Technology, Tsinghua University, Beijing 100084, China*
*[\*xuexx@tsinghua.edu.cn](mailto:xuexx@tsinghua.edu.cn)*

## 1. Impedance matching in radiofrequency circuits

Figure S1a shows a complex load connected to a transmission line. The load is composed of parallel inductance ($L_1$), capacitance ($C_1$), and resistance ($R_L$). The voltage reflection coefficient is given by

$$\Gamma = \frac{U_r}{U_i} = \frac{Z_L - Z_0}{Z_L + Z_0} \tag{S1}$$

where

$$Z_L = \frac{1}{\frac{1}{i\Omega L_1} + i\Omega C_1 + \frac{1}{R_L}} \tag{S2}$$

$\Omega$ is the signal angular frequency; and $Z_0$ is the characteristic impedance of the transmission line. In the narrow-band approximation, $(\Omega - \Omega_1)/\Omega_1 \ll 1$ where $\Omega_1 = 1/\sqrt{L_1 C_1}$ refers to the LC resonant frequency, we have

$$\Gamma \approx -1 + \frac{2B_0}{i(\Omega - \Omega_1) + B_0 + B_L} \tag{S3}$$

where $B_0 = 1/(2Z_0 C_1)$, $B_L = 1/(2R_L C_1)$.

The reflection can only be zero when $\Omega = \Omega_1$ and $B_0 = B_L$. When these conditions are not satisfied, a parallel LC network may be inserted quarter wave apart from the load to eliminate the reflection, as illustrated in Fig. S1b. The reflection coefficient in this case is given by

$$\begin{aligned}
\Gamma &= \frac{U_r}{U_i} = \frac{Z_e - Z_m}{Z_e + Z_m} \\
&\approx -1 + \frac{2Z_L'}{i2Z_L' Z_m C_2 (\Omega - \Omega_2) + Z_m + Z_L'} \\
&= -1 + \frac{2B_m}{i(\Omega - \Omega_2) + B_m + \frac{B_C^2}{i(\Omega - \Omega_1) + B_L}}
\end{aligned} \tag{S4}$$

where $Z_e$ is the complex load seen after the LC matching circuit; $Z_m$ is the characteristic impedance of the new transmission line; $Z_L'$ is the complex load seen after the quarter-wave transmission line, given by $Z_L' = Z_0^2/Z_L$; $\Omega_2 = 1/\sqrt{L_2 C_2}$ the resonant frequency of the LC matching circuit; $B_C = 1/(2Z_0 \sqrt{C_1 C_2})$ and $B_m = 1/(2Z_m C_2)$.

The reflection can be eliminated in the following conditions

$$B_m = \frac{B_C^2 B_L}{(\Omega - \Omega_1)^2 + B_L^2} \tag{S5}$$

$$\Omega - \Omega_2 = \frac{B_C^2 (\Omega - \Omega_1)}{(\Omega - \Omega_1)^2 + B_L^2} \tag{S6}$$

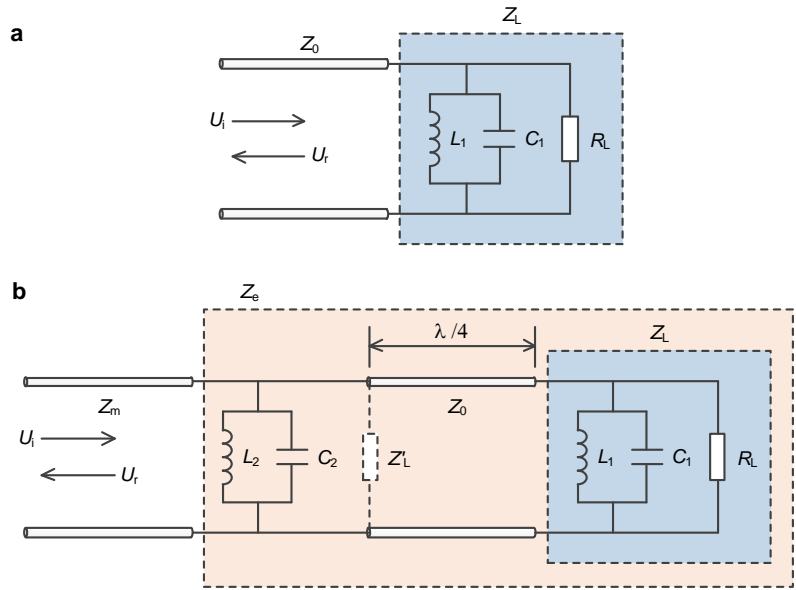

Figure S1 | Impedance matching in radiofrequency circuits. (**a**) A complex load connected to a transmission line. Reflection arises if the signal frequency is not equal to the LC resonant frequency and the load resistance does not match the characteristic impedance of the transmission line. (**b**) A parallel LC network is inserted quarter wave apart from the load to eliminate the reflection.

## 2. Energy transfer in coupled optical cavities

The case of a single cavity pumped by an incident field from port P is illustrated in Fig. S2a. The coupled-mode equations can be written as [S1]

$$\frac{dU_1}{dt} = (i\omega_1 - \kappa_P - \kappa_D) U_1 + \sqrt{2\kappa_P}\, a_P \tag{S7}$$

19 / 30

$$a_T = a_P - \sqrt{2\kappa_P} U_1 \tag{S8}$$

where $U_1$ is the mode amplitude normalized so that $|U_1|^2$ represents the energy in the cavity; $a_P$ and $a_T$ are the field amplitudes at port P and port T, normalized so that $|a_P|^2$ and $|a_T|^2$ represent the power; $\omega_1$ is the angular frequency of the resonator; $\kappa_P$ and $\kappa_D$ are the coupling rates. The amplitude transfer function from port P to port T is given by

$$T(\omega_P) = \frac{a_T}{a_P} = 1 - \frac{2\kappa_P}{i(\omega_P - \omega_1) + \kappa_P + \kappa_D} \tag{S9}$$

where $\omega_P$ is the pump angular frequency. When the pump frequency matches that of the resonator ($\omega_P = \omega_1$) and the coupling rates satisfy $\kappa_P = \kappa_D$, we have $T = 0$ which means the incident wave is completely absorbed by the resonator and vanishes at port T. Based on energy conservation, it is easy to conclude that all the incident power from port P is dissipated to port D. When the pump frequency is detuned from the cavity, $T = 0$ cannot be achieved which means full energy transfer is impossible.

If there is another resonator (cavity 2; assumed intrinsically lossless) coupled to cavity 1 and the pump field, the coupled-mode equations become

$$\frac{dU_1}{dt} = (i\omega_1 - \kappa_D) U_1 + i\kappa_{12} U_2 \tag{S10}$$

$$\frac{dU_2}{dt} = (i\omega_2 - \kappa_P) U_2 + i\kappa_{21} U_1 + \sqrt{2\kappa_P} a_P \tag{S11}$$

$$a_T = a_P - \sqrt{2\kappa_P} U_2 \tag{S12}$$

where $U_2$ and $\omega_2$ are the mode amplitude and angular frequency of cavity 2; the coefficients of coupling between cavity 1 and cavity 2 obey the relation $\kappa_{12} = \kappa_{21}^* = \kappa_C$. The amplitude transfer function from port P to port T is given by

$$T(\omega_P) = \frac{a_T}{a_P} = 1 - \frac{2\kappa_P}{i(\omega_P - \omega_2) + \kappa_P + \dfrac{|\kappa_C|^2}{i(\omega_P - \omega_1) + \kappa_D}} \tag{S13}$$

Complete energy transfer from port P to port D can be achieved with $T = 0$ when the following condition is satisfied

$$\kappa_P = \frac{|\kappa_C|^2 \kappa_D}{(\omega_P - \omega_1)^2 + \kappa_D^2} \tag{S14}$$

$$\omega_P - \omega_2 = \frac{|\kappa_C|^2 (\omega_P - \omega_1)}{(\omega_P - \omega_1)^2 + \kappa_D^2} \tag{S15}$$

Note that Eqs. (S9) and (S13) share the same form as (S3) and (S4) except for a different sign, with the relations $\kappa_P \leftrightarrow B_m$, $\kappa_C \leftrightarrow B_C$, $\kappa_D \leftrightarrow B_L$, $\omega_1 \leftrightarrow \Omega_1$, and $\omega_2 \leftrightarrow \Omega_2$.

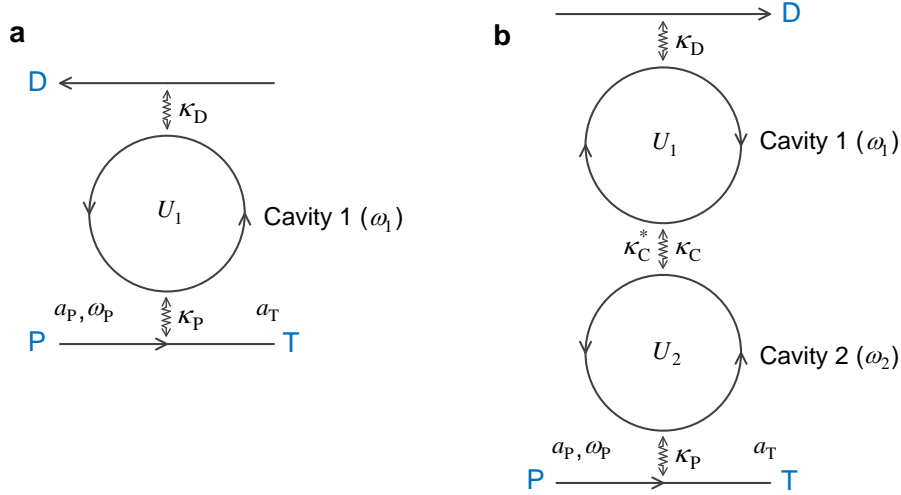

Figure S2 | Energy transfer in optical resonators. (**a**) One single cavity. (**b**) Two mutually coupled cavities.

### 3. Derivation of the coupled nonlinear Schrödinger equations

We fold the dual coupled resonator structure as in Fig S3 so that the propagation of the optical fields in cavity 1 and cavity 2 can be described in the same angular coordinate. The coupled nonlinear Schrödinger equations are written as [S2]

$$\frac{\partial E_1}{R_1 \partial \phi} = \left[ -\alpha_{i1} + ik_1 - k_1' \frac{\partial}{\partial T} - i \frac{k_1''}{2} \frac{\partial^2}{\partial T^2} + i\gamma_1 |E_1|^2 \right] E_1 - \frac{1}{R_1} \sum_{n=-\infty}^{+\infty} \delta(\phi - n2\pi)\left(1 - \sqrt{1-\theta_1}\right) E_1$$
$$+ \frac{1}{R_1} \sum_{n=-\infty}^{+\infty} \delta(\phi - n2\pi - \pi)\left[ i\sqrt{\theta_{12}} E_2 - \left(1 - \sqrt{1-\theta_{12}}\right) E_1 \right]$$
(S16)

$$\frac{\partial E_2}{R_2 \partial \phi} = \left[ -\alpha_{i2} + ik_2 - k_2' \frac{\partial}{\partial T} - i \frac{k_2''}{2} \frac{\partial^2}{\partial T^2} + i\gamma_2 |E_2|^2 \right] E_2 + \frac{1}{R_2} \sum_{n=-\infty}^{+\infty} \delta(\phi - n2\pi)\left[ i\sqrt{\theta_2} E_{in} - \left(1 - \sqrt{1-\theta_2}\right) E_2 \right]$$
$$+ \frac{1}{R_2} \sum_{n=-\infty}^{+\infty} \delta(\phi - n2\pi - \pi)\left[ i\sqrt{\theta_{21}} E_1 - \left(1 - \sqrt{1-\theta_{21}}\right) E_2 \right]$$
(S17)

where $\phi$ is the circular angle; $T$ is the time (usually called fast time in the literature in contrast to a slow time corresponding to the propagation); $R_{1,2}$ radii of MR1 and MR2; $E_{1,2}$ the complex field amplitudes in cavity 1 and cavity 2, normalized so that $|E_{1,2}|^2$ represent the power; $E_{in}$ the pump field amplitude; $\alpha_{i1,i2}$ the intrinsic amplitude loss per unit length; $k_{1,2}$ the wave vectors;

$k'_{1,2} = dk_{1,2}/d\omega\big|_{\omega=\omega_p}$ reciprocal of the group velocities; $k''_{1,2} = d^2k_{1,2}/d\omega^2\big|_{\omega=\omega_p}$ group velocity dispersion; $\gamma_{1,2}$ nonlinear coefficients; $\theta_1$ power coupling ratio between cavity 1 and the drop waveguide; $\theta_2$ power coupling ratio between cavity 2 and the pump waveguide; $\theta_{12} = \theta_{21}$ power coupling ratio between cavity 1 and cavity 2; $\delta(\cdot)$ the Dirac function.

Replace the variables $\phi$, $T$ with $z = R_1\phi$, $\tau = T - R_1\phi k'_1$, the equations become

$$\frac{\partial E_1}{\partial z} = \left[-\alpha_{i1} + ik_1 - i\frac{k''_1}{2}\frac{\partial^2}{\partial \tau^2} + i\gamma_1|E_1|^2\right]E_1 - \sum_{n=-\infty}^{+\infty}\delta(z-nL)\left(1-\sqrt{1-\theta_1}\right)E_1$$
$$+ \sum_{n=-\infty}^{+\infty}\delta\left(z-nL-\frac{L}{2}\right)\left[i\sqrt{\theta_{12}}E_2 - \left(1-\sqrt{1-\theta_{12}}\right)E_1\right] \quad (S18)$$

$$\frac{\partial E_2}{\partial z} = \left[-\alpha_{i2}^s + ik_2^s - \Delta k'^s\frac{\partial}{\partial \tau} - i\frac{k_2''^s}{2}\frac{\partial^2}{\partial \tau^2} + i\gamma_2^s|E_2|^2\right]E_2 + \sum_{n=-\infty}^{+\infty}\delta(z-nL)\left[i\sqrt{\theta_2}E_{in} - \left(1-\sqrt{1-\theta_2}\right)E_2\right]$$
$$+ \sum_{n=-\infty}^{+\infty}\delta\left(z-nL-\frac{L}{2}\right)\left[i\sqrt{\theta_{21}}E_1 - \left(1-\sqrt{1-\theta_{21}}\right)E_2\right] \quad (S19)$$

where $z$ means the propagation distance in cavity 1; $\alpha_{i2}^s$, $k_2^s$, $k_2''^s$, $\gamma_2^s$ are scaled parameters given by $\alpha_{i2}^s = \alpha_{i2}R_2/R_1$, $k_2^s = k_2R_2/R_1$, $k_2''^s = k_2''R_2/R_1$, $\gamma_2^s = \gamma_2R_2/R_1$; $\Delta k'^s$ is the scaled group velocity mismatch given by $\Delta k'^s = k_2'R_2/R_1 - k_1'$; $L = 2\pi R_1$ the round-trip length of cavity 1.

The round-trip phase delays can be written as $k_1L = m_1 2\pi - \delta_{01}$ and $k_2^sL = m_2 2\pi - \delta_{02}$, where $m_1, m_2$ are mode numbers of the cavity modes that are pumped; $\delta_{01}$, $\delta_{02}$ mean the round-trip phase detuning between the pump and the cavity modes. Replace $E_1$ and $E_2$ with $E_1' = E_1 e^{-ik_{m1}z}$ and $E_2' = E_2 e^{-ik_{m2}z}$ where $k_{m1}L = m_1 2\pi$ and $k_{m2}L = m_2 2\pi$, we have

$$\frac{\partial E_1'}{\partial z} = \left[-\alpha_{i1} - i\delta_1 - i\frac{k_1''}{2}\frac{\partial^2}{\partial \tau^2} + i\gamma_1|E_1'|^2\right]E_1' - \sum_{n=-\infty}^{+\infty}\delta(z-nL)\left(1-\sqrt{1-\theta_1}\right)E_1'$$
$$+ \sum_{n=-\infty}^{+\infty}\delta\left(z-nL-\frac{L}{2}\right)\left[i\sqrt{\theta_{12}}E_2'e^{i(m_2-m_1)\pi} - \left(1-\sqrt{1-\theta_{12}}\right)E_1'\right] \quad (S20)$$

$$\frac{\partial E_2'}{\partial z} = \left[-\alpha_{i2}^s - i\delta_2 - \Delta k'^s\frac{\partial}{\partial \tau} - i\frac{k_2''^s}{2}\frac{\partial^2}{\partial \tau^2} + i\gamma_2^s|E_2'|^2\right]E_2' + \sum_{n=-\infty}^{+\infty}\delta(z-nL)\left[i\sqrt{\theta_2}E_{in} - \left(1-\sqrt{1-\theta_2}\right)E_2'\right]$$
$$+ \sum_{n=-\infty}^{+\infty}\delta\left(z-nL-\frac{L}{2}\right)\left[i\sqrt{\theta_{21}}E_1'e^{i(m_1-m_2)\pi} - \left(1-\sqrt{1-\theta_{21}}\right)E_2'\right] \quad (S21)$$

where $\delta_1 = \delta_{01}/L$ and $\delta_2 = \delta_{02}/L$ the averaged phase detuning.

When $|m_1 - m_2|$ is even, the related phase terms in Eqs. (S20) and (S21) are just 1. When $|m_1 - m_2|$ is odd, the phase terms are −1 meaning that there is a $\pi$ phase difference between $E_1'$ and $E_2'$ at the inter-cavity coupling point. In this case, we can do a further variable replacement with $E_1'' = -E_1'$, then we arrive at the same form of equations. To simplify the denotations, we still use $E_1$, $E_2$ to represent $E_1'$ (or $E_1''$), $E_2'$; and $\alpha_{i2}$, $\Delta k'$, $k_2''$, $\gamma_2$ to represent the scaled parameters $\alpha_{i2}^s$, $\Delta k'^s$, $k_2''^s$, $\gamma_2^s$. Finally we have the following set of equations describing the field evolution in cavity 1 and cavity 2, together with those for the fields in the bus waveguides.

$$\frac{\partial E_1}{\partial z} = \left[ -\alpha_{i1} - i\delta_1 - i\frac{k_1''}{2}\frac{\partial^2}{\partial \tau^2} + i\gamma_1 |E_1|^2 \right] E_1 - \sum_{n=-\infty}^{+\infty} \delta(z-nL)\left(1-\sqrt{1-\theta_1}\right)E_1$$
$$+ \sum_{n=-\infty}^{+\infty} \delta\left(z-nL-\frac{L}{2}\right)\left[ i\sqrt{\theta_{12}} E_2 - \left(1-\sqrt{1-\theta_{12}}\right)E_1 \right] \quad (S22)$$

$$\frac{\partial E_2}{\partial z} = \left[ -\alpha_{i2} - i\delta_2 - \Delta k'\frac{\partial}{\partial \tau} - i\frac{k_2''}{2}\frac{\partial^2}{\partial \tau^2} + i\gamma_2 |E_2|^2 \right] E_2 + \sum_{n=-\infty}^{+\infty} \delta(z-nL)\left[ i\sqrt{\theta_2} E_{\text{in}} - \left(1-\sqrt{1-\theta_2}\right)E_2 \right]$$
$$+ \sum_{n=-\infty}^{+\infty} \delta\left(z-nL-\frac{L}{2}\right)\left[ i\sqrt{\theta_{21}} E_1 - \left(1-\sqrt{1-\theta_{21}}\right)E_2 \right] \quad (S23)$$

$$E_{\text{T}} = i\sqrt{\theta_2}\, E_2\big|_{z=nL} + \sqrt{1-\theta_2}\, E_{\text{in}} \quad (S24)$$

$$E_{\text{D}} = i\sqrt{\theta_1}\, E_1\big|_{z=nL} \quad (S25)$$

where $E_{\text{T}}$ and $E_{\text{D}}$ denote the complex field amplitudes at port T and D.

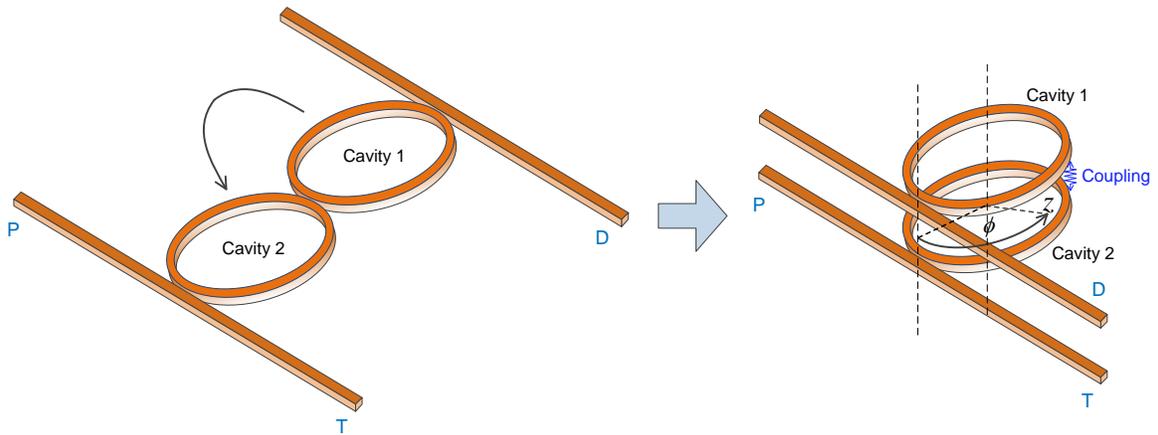

Figure S3 | Illustration of folding the mutually coupled cavities for equation derivation.

## 4. Conditions for maximum conversion efficiency

When the changes of the optical fields in one roundtrip are very small (generally true in microresonator based Kerr comb generation), the discrete terms in Eqs. (S22) and (S23) can be averaged over one roundtrip. We get the following mean-field equations

$$\frac{\partial E_1}{\partial z} = \left[ -(\alpha_{i1} + \alpha_{c1}) - i\delta_1 - i\frac{k_1''}{2}\frac{\partial^2}{\partial \tau^2} + i\gamma_1 |E_1|^2 \right] E_1 + i\kappa_{12} E_2 \tag{S26}$$

$$\frac{\partial E_2}{\partial z} = \left[ -(\alpha_{i2} + \alpha_{c2}) - i\delta_2 - \Delta k' \frac{\partial}{\partial \tau} - i\frac{k_2''}{2}\frac{\partial^2}{\partial \tau^2} + i\gamma_2 |E_2|^2 \right] E_2 + i\kappa_{21} E_1 + i\kappa_p E_{in} \tag{S27}$$

$$E_T = i\kappa_p L E_2 + E_{in} \tag{S28}$$

$$E_D = i\kappa_d L E_1 \tag{S29}$$

where $\alpha_{c1} = \theta_1/(2L)$, $\alpha_{c2} = \theta_2/(2L)$, $\kappa_{12} = \kappa_{21} = \sqrt{\theta_{12}}/L = \sqrt{\theta_{21}}/L$, $\kappa_p = \sqrt{\theta_2}/L$, $\kappa_d = \sqrt{\theta_1}/L$.

When the intracavity fields reach stable states corresponding to comb mode-locking, we have the following equations for the pump mode

$$0 = \left[ -(\alpha_{i1} + \alpha_{c1} + \alpha_{n1}) - i\delta_{1,\text{eff}} \right] E_1 + i\kappa_{12} E_2 \tag{S30}$$

$$0 = \left[ -(\alpha_{i2} + \alpha_{c2}) - i\delta_{2,\text{eff}} \right] E_2 + i\kappa_{21} E_1 + i\kappa_p E_{in} \tag{S31}$$

where $\alpha_{n1}$ is the nonlinear loss due to energy transfer from the pump to the comb, given by $\alpha_{n1} = (\alpha_{i1} + \alpha_{c1}) P_{\text{comb}}/P_{\text{pump}}$; $P_{\text{comb}}$ is the power of the comb lines excluding the pump in cavity 1; $P_{\text{pump}}$ is the power of the pump in cavity 1; $\delta_{1,\text{eff}}$ and $\delta_{2,\text{eff}}$ are the effective detuning. Note that here we have omitted the mode coupling between the two cavities around mode-crossing regions which transfers a small portion of comb power from cavity 1 to cavity 2. In the following analysis in this section, we will use $E_1$ and $E_2$ to represent only the amplitude of the pump. The pump field in cavity 2 can be obtained by solving Eqs. (S30) and (S31). The result is

$$E_2 = \frac{i\kappa_p E_{in}}{\alpha_{2t} + i\delta_{2t}} \tag{S32}$$

where $\alpha_{2t} = \alpha_{i2} + \alpha_{c2} + \alpha_{21}$; $\delta_{2t} = \delta_{2,\text{eff}} + \delta_{21}$; $\alpha_{21}$ and $\delta_{21}$ represent loss and detuning induced by coupling with cavity 1, determined by

$$\alpha_{21} + i\delta_{21} = \frac{|\kappa_{12}|^2}{\alpha_{1,\text{eff}} + i\delta_{1,\text{eff}}} \tag{S33}$$

where $\alpha_{1,\text{eff}}$ means the total effective loss of the pump in cavity 1, given by $\alpha_{1,\text{eff}} = \alpha_{i1} + \alpha_{c1} + \alpha_{n1}$. The pump field at port T is then

$$E_T = \left(1 - \frac{\kappa_p^2 L}{\alpha_{2t} + i\delta_{2t}}\right) E_{\text{in}} \tag{S34}$$

For maximum conversion efficiency, the pump should be strongly depleted with $E_T = 0$, corresponding to the condition

$$\delta_{2t} = 0 \tag{S35}$$

$$\alpha_{2t} = \kappa_p^2 L \tag{S36}$$

The external conversion efficiency, defined as the ratio of the pump at port T transferred to the comb at port D excluding the pump, is approximated by

$$\eta_e \approx \frac{\alpha_{21}}{\alpha_{21} + \alpha_{i2}} \cdot \frac{\alpha_{c1}}{\alpha_{c1} + \alpha_{i1}} \cdot \eta_i \tag{S37}$$

The first term represents the ratio of the inter-cavity coupling induced effective loss over the total transfer loss in cavity 2 (note that the transfer loss here refers to the energy transferred from the pump to other forms (comb or heat) or out of the coupled cavity system (e.g., due to scattering); the waveguide coupling loss (i.e., $\alpha_{c2}$) does not constitute a transfer loss because the energy is not really lost if we consider the whole system including the cavity and the bus waveguide); the second term represents the ratio of the dropping loss over the total loss in cavity 1; the last term means the internal conversion efficiency in cavity 1 defined as the ratio of the comb power excluding the pump over the total power. For the results shown in Fig. 5 of the main paper, the first and second terms are 1 since the resonators are assumed lossless. The external efficiency according to Eq. (S37) is equal to the internal efficiency which is 96.1% − quite close to the numerical result (94.2%). The deviation is attributed to the mode crossing effect that we have omitted. For the results shown in Fig. S7 in the following, the estimated external efficiency is 83.3% which is again close to the numerical result (81.6%).

## 5. Thermo-optic effect of fiber ring cavities

In our experiments, we observed red shifting of the fiber ring cavities induced by thermo-optic effect. Figure S4 shows the transmission curves of cavity 2 measured in the backward direction. The voltage applied on fiber stretcher 2 is scanned with different speeds. When the scanning speed is low, the transmission is deformed to a triangular shape which can be frequently observed in microresonator based frequency comb generation. The thermal response time is estimated to be in a sub-second level.

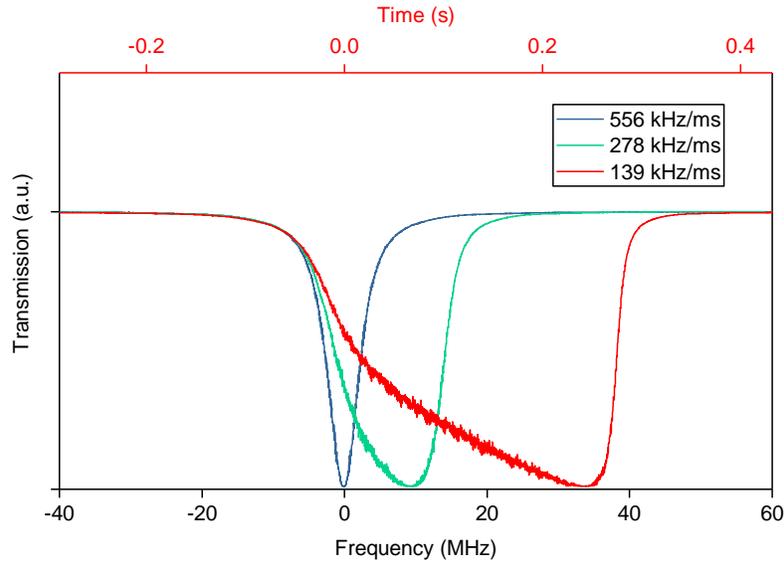

Figure S4 | Transmission of cavity 2 measured with different scanning speeds. The bottom horizontal axis means the frequency detuning of the cavity relative to the pump laser. The top axis shows the time scale when the scanning speed is 139 kHz/ms.

## 6. Simulated high-resolution spectra of fiber cavity solitons

When there is one single soliton in cavity 1, the simulated comb spectra in cavity 1 and cavity 2 are shown in Figs. S5. The fine spectral features that are not revealed in Fig. 3 of the main paper can be observed.

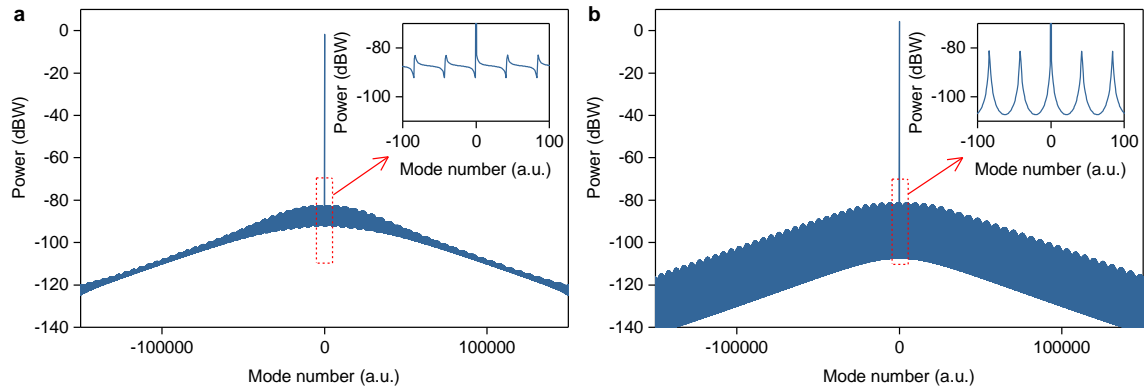

Figure S5 | Numerically simulated comb spectra in (**a**) cavity 1 and (**b**) cavity 2.

## 7. Numerical simulation of mutually coupled optical microresonators

Here, we present more numerical results by considering microresonators with an intrinsic Q of $10^7$. Such Q levels have been demonstrated with the state-of-the-art integration techniques [S3]-[S5]. The energy of the soliton circulating in cavity 1 is dissipated through intrinsic loss and coupling loss. Higher conversion efficiency can be expected if the ratio of the coupling loss over the intrinsic loss is higher, as explained by Eq. (S37). Therefore, we assume cavity 1 is over-coupled to the drop waveguide corresponding to an external Q of $0.5 \times 10^6$. The power coupling ratio between cavity 2 and the pump waveguide is optimized for high conversion efficiency and also results in an over-coupling condition. The simulation parameters are as follows $FSR_1 = 200$ GHz, $FSR_2 = 192.6$ GHz, $L = 1.499$ mm, $\alpha_{i1} = \alpha_{i2} = 0.2027$ m$^{-1}$, $k_1'' = -100$ ps$^2$km$^{-1}$, $k_2'' = 100$ ps$^2$km$^{-1}$, $\gamma_1 = \gamma_2 = 1$ m$^{-1}$W$^{-1}$, $\Delta k' = 1.282 \times 10^{-10}$ s·m$^{-1}$, $\theta_1 = 8.507 \times 10^{-3}$, $\theta_2 = 1.155 \times 10^{-2}$, $\theta_{12} = 3.038 \times 10^{-3}$.

As demonstrated above, the soliton can be stimulated by a single-shot pulse trigger. Another way to obtain the soliton is spontaneous excitation with modulational instability as has been widely employed in single-resonator comb generation. To do so, the CW pump laser is tuned into the microresonator from the blue side (i.e., pump laser wavelength shorter than the resonant wavelength). In our case of mutually coupled microresonators, one prerequisite of obtaining modulational instability in cavity 1 is getting enough power build-up in its cavity. Since cavity 1 is not directly coupled to the pump waveguide, the resonance of cavity 1 is only visible through its coupling to cavity 2. To evaluate the power build-up condition, it is helpful to look at the linear transmission of the mutually coupled microresonators.

**Linear transmission.** Figure S6a shows the linear transmission of port D and port T. The input probe field is injected from port P. Exact frequency matching is assumed for the modes of cavity 1 and cavity 2 which are pumped in comb generation (corresponding to zero frequency in the figure). The modes become hybridized due to inter-cavity coupling. The transmission at port T shows a high extinction ratio close to critical coupling (see the region marked with a dash box). Strong field build-up in cavity 1 can be observed from the transmission of port D. For the modes other than the pumped, the frequency detuning between cavity 1 and cavity 2 gets larger. Their resonances tested at port T tend to the natural transmission of cavity 2 (i.e. the case of no inter-cavity coupling) and are much weaker (recall that cavity 2 is over-coupled to the bus waveguide; note that in a frequency range larger than that shown in Fig. S6a, the modes of cavity 1 and cavity 2 may become closer again periodically). Therefore, to achieve strong field build-up, the pumped modes of cavity 1 and cavity 2 should be close enough in the comb initiation stage. Figure S6b shows zoom-in plots of the pumped mode with different detuning between cavity 1 and cavity 2, which is a typical picture of avoided mode crossing. Experimentally measured results as Fig. S6b can be found in [S6].

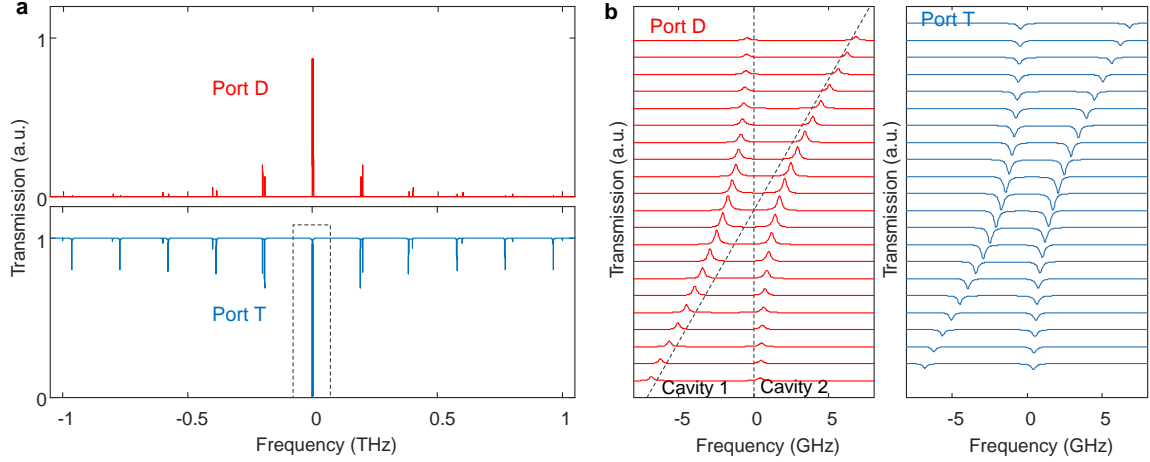

Figure S6 | Linear transmission of mutually coupled microcavities. The probe field is injected from port P. (**a**) Transmission of drop port (port D) and through port (port T). The modes around zero frequency (marked in a dash box) are pumped in comb generation. The natural frequencies of cavity 1 and cavity 2 are assumed equal for the pumped modes. (**b**) Zoom-in of the pumped modes with different detunings between cavity 1 and cavity 2, showing a typical picture of avoided mode crossing. The dash lines indicate the natural resonant frequencies.

**Spontaneous soliton excitation from chaos.** Figure S7 shows one example of spontaneous soliton excitation. In stage I, the wavelength of cavity 1 is slightly red detuned with respect to cavity 2. The pump laser scans across the resonance from the red side as in usual microcomb generation experiments [S7] (see the illustration in Fig. S7c, left). This is done in numerical simulations by synchronously changing the frequency detuning of cavity 1 and cavity 2 with respect to the pump $(\Delta f_1, \Delta f_2)$ from $(-1.910\,\text{GHz}, -2.698\,\text{GHz})$ to $(1.273\,\text{GHz}, 0.4852\,\text{GHz})$ at a speed of $12.73\,\text{GHz}/\mu\text{s}$. The pump power is 500 mW. The evolution of the optical field in cavity 1 is shown in Fig. S7a. The power transmission at port T and port D is shown in Fig. S7b. As the pump is tuned into the resonance, modulational instability occurs and frequency combs are generated. The comb then turns to a chaotic state with randomly changing time-domain waveforms. As the pump laser is tuned further, power transition is observed and a single soliton is generated after transition. In stage II, cavity 1 is tuned to shorter wavelength while the wavelengths of the pump laser and cavity 2 are fixed (see the illustration in Fig. S7c, right). This corresponds to increasing the detuning between cavity 1 and the pump ($\Delta f_1$) from $1.273\,\text{GHz}$ to $4.775\,\text{GHz}$ at a speed of $12.73\,\text{GHz}/\mu\text{s}$. The soliton is compressed and the peak power increases in this process. In the meanwhile, the pump power is reduced adaptively to suppress soliton breathing. In stage III, the frequency detuning of cavity 1 and cavity 2 is fixed at $4.775\,\text{GHz}$ and $0.4852\,\text{GHz}$, respectively. A stable and high-efficiency single soliton is finally obtained with a pump power of 20 mW. The soliton FWHM is 26 fs. The time-domain waveforms and the comb spectra at different times in the soliton excitation process are shown in Fig. S7d. In stage III, the power from port D and port T is 86.6% and 1.8% respectively. About 11.6% of the input power is lost due to the intrinsic losses. The pump-to-comb conversion efficiency is 81.6%.

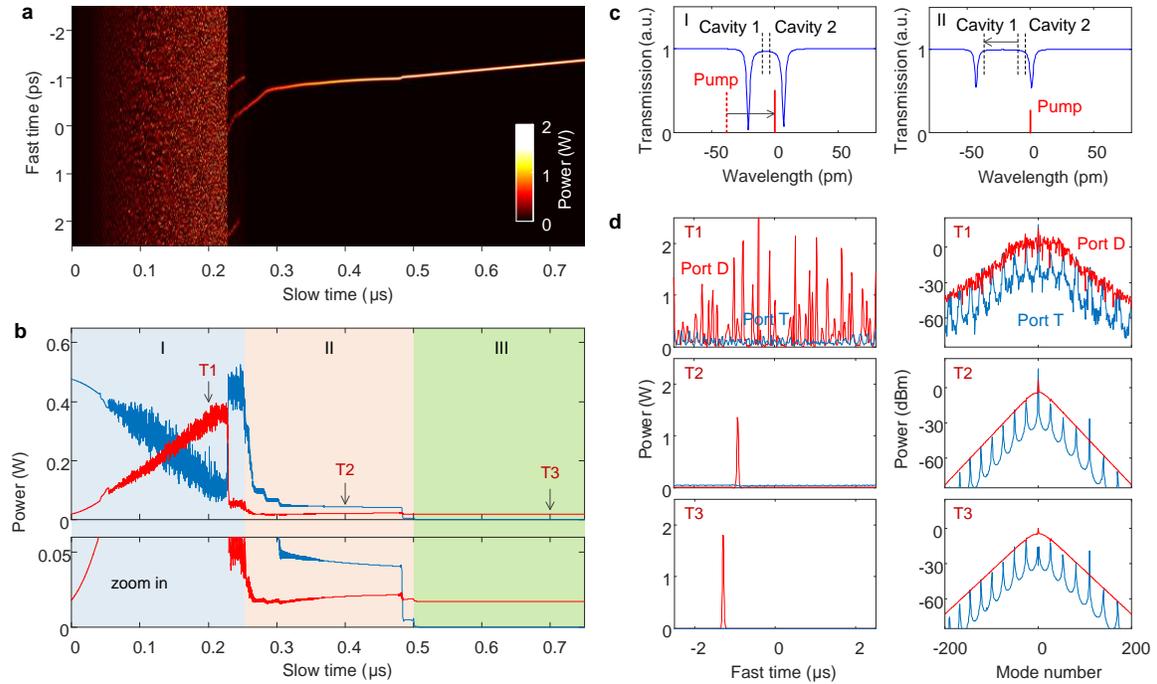

Figure S7 | Spontaneous soliton excitation from chaos. (**a**) Evolution of the time-domain waveform in cavity 1. The fast time is related to the angular position in the cavity. The slow time is related to the propagation distance. (**b**) Power transmission at port T and port D. In stage I, the pump laser is tuned from shorter wavelength to longer wavelength with a power of 500 mW. A single soliton is obtained in cavity 1 at the end of laser tuning. In stage II, cavity 1 is tuned to shorter wavelength, while the wavelengths of the pump laser and cavity 2 are fixed and the pump power is adaptively reduced to suppress soliton breathing. In stage III, the optical field in cavity 1 evolves to a stable high-efficiency soliton. The pump power is 20 mW. (**c**) Linear transmission in stage I (left), and at the end of stage II and in stage III (right). The dash lines show the natural frequencies of cavity 1 and cavity 2. (**d**) Time-domain waveforms (left) and comb spectra (right) at port T and port D at different times marked in (b).

**Soliton regulation induced by mode coupling.** Interestingly, the coupling of cavity 1 and cavity 2 not only enables high energy conversion efficiency, but also affects the soliton dynamics in cavity 1. We investigated the repeatability of spontaneous soliton excitation described above, and found that the soliton transition from chaos is more deterministic than in a single microresonator without mode coupling. Figure S8 shows the overlapped power transmission traces at port D, plotted based on the results of 100 numerical tests. The different colors represent different probabilities. About 70% of the tests transition to a single-soliton state while 30% transition to a CW state with no solitons. No multi-soliton states are observed. We attribute the more deterministic single-soliton transition to a similar mechanism as in a single microresonator with spatial mode coupling [S8] or fundamental-second-harmonic coupling [S9]. Moreover, soliton drifting can be observed from Fig. S7a in stage III, which implies a deviation of the soliton repetition rate from the FSR of cavity 1. The retrieved soliton repetition rate is 200.0013 GHz in comparison to the 200-GHz FSR of cavity 1. This effect is also attributed to the coupling between cavity 1 and cavity 2, and was again observed in a single microresonator with mode coupling [S9], [S10].

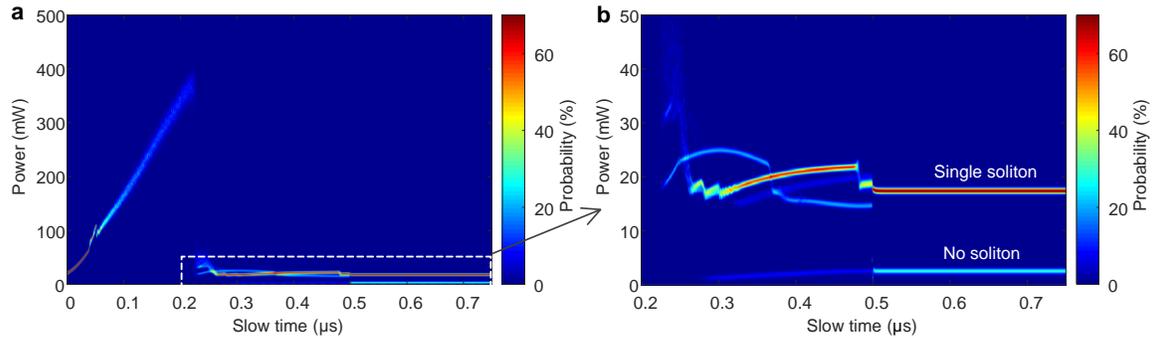

Figure S8 | Repeatability of spontaneous soliton excitation. (**a**) Overlapped power transmission traces at port D. The different colors indicate different probabilities which are calculated based on results of 100 numerical tests. (**b**) Zoom-in of the dash lined region in (a).